\documentclass[showpacs,twocolumn,floats]{revtex4}
\usepackage{graphicx}
\usepackage{amsmath}
\usepackage{psfrag}
\begin{document}
\title{Cross-correlation of two interacting conductors}
\author{M. C. Goorden and M. B\"uttiker}
\affiliation{D\'epartement de Physique Th\'eorique, Universit\'e de Gen\`eve,
  CH-1211 Gen\`eve 4, Switzerland.}
\date{May 27, 2008}
\begin{abstract}
We calculate the current cross-correlation for two weakly interacting mesoscopic conductors. Our derivation is based on the two-particle scattering matrix derived in Goorden and B\"uttiker [Phys. Rev. Lett. {\bf 99}, 146801 (2007)]. We include the Fermi sea in the leads into the theory and show how to calculate transport quantities and specifically cross-correlations. We focus on the zero-frequency current cross-correlation of two chaotic quantum dots and calculate the magnitude of its fluctuations with the help of Random Matrix Theory.
\end{abstract}

\maketitle

{\begin{figure}[t]
\begin{center}
\psfrag{ald}{$a_L^\dag$}
\psfrag{bld}{$b_L^\dag$}
\psfrag{dld}{$d_L^\dag$}
\psfrag{cld}{$c_L^\dag$}
\psfrag{ard}{$a_R^\dag$}
\psfrag{brd}{$b_R^\dag$}
\psfrag{drd}{$d_R^\dag$}
\psfrag{crd}{$c_R^\dag$}
\psfrag{1}{$\hat{Q}^{\rm I}$}
\psfrag{2}{$\hat{Q}^{\rm II}$}
\psfrag{l12}{$\lambda \hat{Q}^{\rm I}\hat{Q}^{\rm II}$}
\psfrag{NL1}{$N_L^{\rm I}$}
\psfrag{NL2}{$N_L^{\rm II}$}
\psfrag{NR1}{$N_R^{\rm I}$}
\psfrag{NR2}{$N_R^{\rm II}$}
\psfrag{V1}{$V^{\rm I}$}
\psfrag{V2}{$V^{\rm II}$}
\includegraphics[width=8cm]{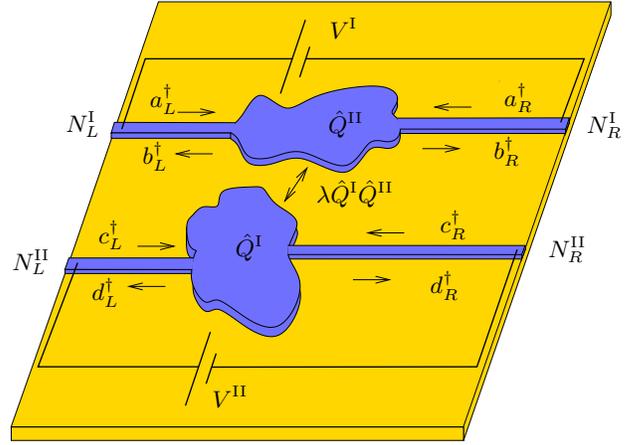}
\end{center}
\caption{(color online) Two quantum dots coupled via an interaction
$\lambda\hat{Q}^{\rm I}\hat{Q}^{\rm II}$. The operators $a^\dag$ ($c^\dag$) create
incoming electrons in the scattering states in the left (L) and right (R) leads of dot ${\rm I}$ (${\rm II}$), while
  $b^\dag$ ($d^\dag$) are similar operators for outgoing electrons. The dots
are biased by voltages $V^{\rm I/II}$.
 The leads
have $N_{L/R}^{\rm
  I/II}$ channels.}
\label{dotsfig}
\end{figure}}
In recent work \cite{Goo07} we analyzed scattering of two electrons moving each in a separate conductor. The two conductors are in proximity of each other (see Fig. 1). There is no carrier exchange between the conductors but possibly interaction through the long-range Coulomb force. We assume that carriers in the leads are perfectly screened and thus in the  leads they are effectively non-interacting.   However, when both electrons are simultaneously
inside the conductors, both conductors might have a net charge affecting
electron motion in the other nearby conductor. The presence of an excess electron in one dot changes the scattering properties of the electron in the other. This interaction correlates the two electrons. In our recent work we presented a discussion of the resulting correlation in terms of a  two-particle scattering matrix. In the absence of interaction the two-particle scattering matrix is just a product of single-particle scattering matrices. Weak interaction adds a term to this matrix proportional to the interaction strength and proportional to the product of the density of states matrices of the two conductors \cite{Goo07}.

The physical observable of interest is the current cross-correlation between the two conductors. For two parallel conductors in the Coulomb blockade regime this correlation has recently been measured \cite{McC07}. Below we present a calculation of the current cross-correlation for dots which are connected to leads via quantum point contacts with several open channels. In such a case, both the Coulomb interaction within each dot as well as the Coulomb interaction between the electrons in the two different dots is effectively weak \cite{Bro05}. It is then sufficient to find the current cross-correlation to leading order in the interaction strength. Our formulation permits us to treat quantum interference effects in each conductor using Random Matrix Theory.

However, the two-particle scattering states are not sufficient to find the current cross-correlation between the two conductors. The incoming state in a transport experiment is fundamentally a many particle state. In the zero temperature limit each incident state with an energy in the transport window is occupied with probability one. Hence it is necessary to demonstrate how a two-particle scattering matrix is nevertheless an useful object. In this work we extend our earlier discussion \cite{Goo07} by including the Fermi sea.

Configurations which can be viewed as two parallel and separate conductors are of interest in a number of experiments in addition to Ref. \cite{McC07}. We mention only briefly the problem of Coulomb drag \cite{Mor01,Yama}. Of special interest are geometries in the quantum Hall regime where separately contacted edge states might be viewed as separate conductors. In adiabatic geometries, over large distances, edge states do not exchange carriers. Examples are quantum dots where an inner interfering edge state is dephased by an outer noisy edge channel \cite{moty1} or a Mach-Zehnder interferometer \cite{Ji03,litv,roche1,roche2,roche3,litv2,See,Chu} which interacts with a current carrying edge channel \cite{moty3,Ned07}. It is important to note that 
these arrangements differ from the more widely discussed quantum  measurement problem where a conductor, a quantum point contact or a quantum dot, is used to measure a (charge) qubit since in the problems of interest here both conductors carry current. Nevertheless also here one conductor can be viewed as the detector testing the {\it transport state} of the other conductor. For some problems, especially if screening in the two conductors is poor, it might be necessary to go beyond the weak interaction limit and beyond an approach based on two-particle collisions. Indeed some works \cite{moty3,Ned07,ned08,sim07,sukh08} investigate many particle scattering processes to account for the experimental observations. 

In the future it will certainly be possible to inject single carriers into scattering states and guide them toward a region where they interact. The two-particle scattering matrix would then be the obvious object to describe such a scattering experiment.  Single-particle injection can be achieved by replacing electron reservoirs by single electron devices (pumps) in the Coulomb blockade regime or by mesoscopic capacitors \cite{gabe,nigg} subject to large amplitude pulses \cite{feve,moska}. The latter have the advantage that carriers
with well-defined energy are injected.

Our interest is in two-particle effects with the two particles located in separate conductors, but two-particle processes within one conductor have also been investigated \cite{qing03}. In this case, even in the absence of interactions, indistinguishability of carriers leads to exchange interference effects in correlations \cite{PRL92}. In particular in non-interacting systems shot noise is a probe of two-particle physics \cite{PRL92}. Indeed there exist geometries which permit an explicit demonstration of statistical two-particle correlations: a geometry in which conductance exhibits no Aharonov-Bohm effect but current correlations exhibit a two-particle Aharonov-Bohm effect was proposed and analyzed in Ref. \cite{Sam04} and recently experimentally demonstrated in Ref. \cite{Ned07exp}. In a non-interacting system a correlated N-particle scattering event requires the observation of a N-th (or higher) order cumulant of a current correlation \cite{LL93,Sim05}. In contrast such a conceptually appealing hierarchy does not exist in interacting systems. For some time, the properties of interacting two-particle states have been of interest in disordered and chaotic systems  \cite{Jac97}. Here it seems possible that at least for certain scatterers (isolated dots) one can go beyond the two-particle, weak interaction approach and discuss an N-particle scattering matrix \cite{Leb07,dhar08}. However these discussions \cite{Jac97,Leb07} do not include a Fermi see. In impurity problems \cite{Sel06,moty4,hur7} where one deals with a Fermi sea, it is often not possible to give scattering states explicitly.  

Below we present the model and give the main results for the two-particle
scattering matrix from our previous work (Section I). Inclusion of the Fermi
sea is discussed in Section II and a general expression for cross-correlations
is derived in Section III. In Section IV we focus on the zero-frequency current
cross-correlation and in Section V we investigate the case of two cavities with many channel contacts and evaluate the cross-correlation with the help of Random Matrix Theory.

\section{Model}

The type of system we consider is depicted in Fig.\ \ref{dotsfig}. Two
scatterers are both coupled to two non-interacting leads. The
derivation in this paper is valid for any scatterer, but we will refer
to them as quantum dots.
The first (second) system has
$N_L^{\rm I}$ ($N_L^{\rm II}$) channels in the left lead , $N_R^{\rm I}$ ($N_R^{\rm II}$) channels in the right lead
and the sum is denoted by $N^{\rm I/II}=N_L^{\rm
  I/II}+N_R^{\rm I/II}$. 
The dots are biased with voltages $V_L^{\rm I/II}/V_R^{\rm I/II}$ in the
left/right leads. We define voltage differences $ V^{\rm
  I/II}=V_L^{\rm I/II}-V_R^{\rm I/II}$.
We denote by $a_i^\dag(E)$ ($b_i^\dag(E)$) the creation operator for electrons
in the ingoing (outgoing) scattering state in channel $i$ with energy $E$. 
The index $i \in 1\dots N^{\rm I}_L$ refers to a channel in the left lead, while
$i\in N^{\rm I}_L+1\dots N^{\rm I}$ signifies that  the channel
is in the right lead.
The equivalent operators for the second dot are denoted by  $c_i^\dag(E)$
($d_i^\dag(E)$). Ingoing and outgoing carriers of the non-interacting dots are
related by the single-particle scattering matrices; $b_i(E)=\sum_jS_{ij}^{\rm
  I}(E)a_j(E)$ and $d_i(E)=\sum_jS_{ij}^{\rm II}(E)c_j(E)$. We assume a
coupling between dots of the form ${\lambda} \hat{Q}^{\rm I}\hat{Q}^{\rm II}/e^2$.
Here $\lambda$ is a coupling energy and $\hat{Q}^{\rm I/II}$ are the total charge operators on the dots.

The two-particle scattering matrix is the starting point of this paper and we will therefore first summarize its properties.
Although the derivation in Ref.\ \cite{Goo07} is for single-channel leads, it is straightforward to extend it and to show that
the final result is also valid for the multi-channel leads considered in this paper.
Incoming and outgoing two-particle states are related via the two-particle scattering matrix $\delta S$,
\begin{align}
\label{invsout}
&b_i(E_1)d_j(E_2)=\sum_{kl}[S_{ik}^{\rm I}(E_1)S_{jl}^{\rm
  II}(E_2)a_k(E_1)c_l(E_2)+
\nonumber\\
&\int d\epsilon\delta S_{ik,jl}(E_1,E_2,E_1+\epsilon,E_2-\epsilon) a_k(E_1+\epsilon) c_l(E_2-\epsilon)].
\end{align}
The first part of this equation describes the scattering in the non-interacting
dots and the second part captures the effects of the interaction. While
the total energy of the particles is conserved in the scattering process,
there can be an
exchange of energy $\epsilon$. Up to first order in the coupling energy
$\lambda$, the two-particle scattering matrix depends on the non-interacting scattering matrices via the relation
\begin{align}
\label{deltaS}
\delta S(E_1,E_2,E_3,E_4)=&-
2\pi i\lambda  S^{\rm I}(E_1){\cal N}^{\rm I}(E_1,E_3)\otimes\nonumber\\
& S^{\rm II}(E_2){\cal N}^{\rm II}(E_2,E_4).
\end{align}
In this notation the close connection with the density of states
matrix \cite{Ped98,But99}
\begin{align}
\label{Ndens}
{\cal N}^{\rm I}(E,E')=S^{\rm I\dag}(E)\frac{S^{\rm
    I}(E)-S^{\rm I}(E')}{2\pi i(E-E')},
\end{align}
is apparent.
The diagonal element ${\cal N}_{ii}$ of this matrix is the part of the density of
states associated with incoming carriers from channel $i$. Charge fluctuations of
non-interacting dots at
frequency $E-E'$ can be described by its off-diagonal elements. In the limit $E\rightarrow
E'$ it reduces to the famous Wigner-Smith delay time matrix \cite{Smi60,Bro97}.
For later use we also give the definition of the closely related charge fluctuation operator,
\begin{align}
\label{Nop}
\hat{\cal N}^{\rm I}(\epsilon)=\int dE
\sum_{ij}{\cal N}^{\rm I}_{ij}(E+\epsilon,E)a^\dag_i(E+\epsilon) a_j(E).
\end{align}
The density of states matrix and charge fluctuation operator for
dot ${\rm II}$ can be defined in an equivalent manner.

\section{Outgoing wave function}

As already mentioned in the introduction the incoming state in a transport experiment is not a two-particle state but contains many particles. The incoming multi-particle state will scatter into an outgoing
multi-particle state. The relation between multi-particle and two-particle
scattering for non-relativistic particles in a two-particle potential has been addressed in Ref.\
\cite{Ros65}, where multi-particle scattering amplitudes are expressed
in terms of the two-particle scattering matrix. In first order in the
interaction, only the terms with two particles interacting, while the other
particles are unaffected by the interaction, survive (the so-called
disconnected diagrams).
We will use this result to determine the outgoing wave function.

Our
approach has many elements in common with the redefinition of the vacuum
\cite{Sam03,Sam05} used in the context of quasi particle entanglement,
but in contrast to these works we do not assume that we are in the tunneling limit.
At zero temperature the wave function of the leads in terms of incoming
operators $a^\dag$ and $c^\dag$ is given by
\begin{equation}
\label{psiin}
|\psi\rangle=\prod_{m=1}^{N^{\rm
 I}}\prod_{E = E_{m}^{\rm I} + eV_m^{\rm I}}^{E_F + eV_m^{\rm I}}a^\dag_{m}(E)|0\rangle\otimes\prod_{n=1}^{N^{\rm II}}\prod_{E= E_{n}^{\rm II}+ eV_n^{\rm II}}^{E_F+eV_n^{\rm
 II}}c^\dag_{n}(E)|0\rangle.
\end{equation}
We assume equal Fermi energies
$E_F$ in all leads, while 
\begin{align}
V_m^{\rm I/II}=\left\{
\begin{array}{cc}
V_L^{\rm I/II}&\mbox{for $1\le m\le N_L^{\rm I/II}$,}\\
V_R^{\rm I/II}&\mbox{else,}
\end{array}\right.
\end{align}
is the bias voltage and $E_m^{\rm I/II}$ the channel threshold in channel $m$  
of dot ${\rm I/II}$.
We can rewrite the wave function in terms of the outgoing $b^\dag$ and $d^\dag$ operators as
\begin{equation}
\label{psiout}
|\psi^{\rm out}\rangle=|\psi^{\rm I}\psi^{\rm II}\rangle+|\delta\psi\rangle.
\end{equation}
The first part is the non-interacting outgoing state,
\begin{align}
\label{defpsiout}
&|\psi^{\rm I/II}\rangle=\prod_{m=1}^{N^{\rm I/II}}\prod_{E = E_{m}^{\rm I/II} + eV_{m}^{\rm I/II}}^{E_F + eV_m^{\rm
  I/II}}\phi^{\rm
  I/II}_{m}(E)|0\rangle, \\
\label{defphi}
&\phi_{k}^{\rm I}(E)=\sum_{j}S^{\rm I}_{jk}(E) b_{j}^\dag(E),
\,\phi_{k}^{\rm II}(E)=\sum_{j}S^{\rm II}_{jk}(E) d_{j}^\dag(E).
\end{align}
The interacting part is given by
\begin{equation}
\label{deltapsi}
|\delta\psi\rangle=\sum_{ij}\int dE_1dE_2|\delta\psi_{iE_1,jE_2}\rangle.
\end{equation}
We have defined $|\delta\psi_{iE_1,jE_2}\rangle$ to be the change in
outgoing wave function if an incoming particle with energy $E_1$ in channel $i$ of
dot ${\rm I}$ interacts with an incoming particle with energy $E_2$ in channel $j$ of dot
${\rm II}$, while all the other particles are unaffected. Eq.\ (\ref{deltapsi})
expresses the fact that the outgoing state is a combination of all possible
two-particle processes. 

Let us proceed to calculate $|\delta\psi_{iE_1,jE_2}\rangle$. We write
\begin{align}
\label{deltapsiij}
|\delta&\psi_{iE_1,jE_2}\rangle=\int d \epsilon\sum_{k,l}\delta S_{ki,lj}
 (E_1+\epsilon,E_2-\epsilon,E_1,E_2)\nonumber\\
&|\psi_{k,E_1+\epsilon}^{'\rm I}\rangle \otimes|\psi_{l,E_2-\epsilon}^{'\rm II}\rangle.
\end{align}
The state $|\psi_{k,E_1+\epsilon}^{'\rm I}\rangle$ consists, like $|\psi^{\rm I}\rangle$ defined in Eq.\ (\ref{defpsiout}), of a product (over energy $E$ and channel index $m$) of creation operators $\phi^{\rm I}_m(E)\equiv \sum_jS_{jm}^{\rm I}(E)b_j^\dag(E)$. In fact it is almost equal to $|\psi^{\rm I}\rangle$, except for one term in the product; one should make the substitution
$\phi^{\rm I}_k(E_1)\rightarrow b^\dag_k(E_1+\epsilon)$.
We can therefore write
\begin{align}
|\psi_{k,E_1+\epsilon}^{'\rm I}\rangle&=\nonumber\\
&\phi_1^{\rm I}(eV_L^{\rm I})\dots b^\dag_{k}(E_1+\epsilon)\dots\phi_{N^{\rm
 I}}^{\rm I}(E_F+eV_R^{\rm I})|0\rangle.
\end{align}
Similarly, $|\psi_{l,E_2-\epsilon}^{'\rm II}\rangle$ is almost equal to $|\psi^{\rm II}\rangle$, except for the term $\phi_l^{\rm II}(E_2)$ and therefore
\begin{align}
|\psi_{l,E_2-\epsilon}^{'\rm II}\rangle&=\nonumber\\
&\phi_1^{\rm II}(eV_L^{\rm II})\dots d^\dag_{l }(E_2-\epsilon)\dots\phi_{N^{\rm II}}^{\rm II}(E_F+eV_R^{\rm II})|0\rangle.
\end{align}
So in this
notation the incoming particles in channels $i$ and $j$ have scattered into
channels $k$ and $l$ exchanging an energy $\epsilon$. The amplitude of this
process is given by $\delta S_{ki,lj}
 (E_1+\epsilon,E_2-\epsilon,E_1,E_2)$. The summation over all possible outgoing
channels and energies is necessary to capture all possible scattering events.

We use the fact that $\phi^{\rm I\dag}_i(E_1)\phi^{\rm I}_i(E_1)|0\rangle=|0\rangle$ and we insert
$\phi_i^{\rm I\dag}(E_1)\phi^{\rm I}_i(E_1)$ in front of the first $|0\rangle$ in Eq.\
(\ref{deltapsiij}). Using the Fermionic anti-commutation rules, one can easily
show that
$\phi^{\rm I\dag}_i(E_2)\phi^{\rm I}_i(E_2)$ commutes with $\phi^{\rm I}_j(E)$
for $i\ne j$, while $b^\dag_{k}(E_1+\epsilon)\phi^{\rm
  I\dag}_i(E_1)$ commutes with $\phi^{\rm I}_j(E)$ for $i\ne j$ or
$E_1\ne E$. Using these properties we can reorder our operators.
Similarly we insert $\phi_j^{\rm II\dag}(E_1)\phi^{\rm II}_j(E_1)$ in
front of the second $|0\rangle$ in Eq.\ (\ref{deltapsiij}) and after
reordering we find
\begin{align}
\label{deltapsiij2}
|\delta\psi_{iE_1,jE_2}\rangle=\int d\epsilon\sum_{k,l}\delta S_{ki,lj}
 (E_1+\epsilon,E_2-\epsilon,E_1,E_2)&\nonumber\\
b^\dag_{k}(E_1+\epsilon)  \phi^{\rm
 I\dag}_{i}(E_1)d^\dag_{l }(E_2-\epsilon) \phi^{\rm II\dag}_{j}(E_2)\,\,\,|\psi^{\rm
 I}\psi^{\rm II}\rangle&\nonumber\\
=-2\pi i\lambda\int d\epsilon\sum_{kl} {\cal N}_{ki}^{\rm
 I}(E_1+\epsilon,E_1){\cal N}_{lj}^{\rm
 II}(E_2-\epsilon,E_2)&\nonumber\\
 a_k^\dag(E_1+\epsilon) a_i(E_1)c^\dag_l(E_2-\epsilon)c_j(E_2)|\psi^{\rm
 I}\psi^{\rm II}\rangle.&
\end{align}
 To obtain the
last equality we have used Eqs.\ (\ref{deltaS}) and
(\ref{defphi}) and the single-particle scattering matrices to
rewrite everything in terms of ingoing operators $a^\dag$ and $c^\dag$ (this is allowed because we are only
interested in terms linear in the interaction).
Combining Eqs.\ (\ref{deltapsi}) and (\ref{deltapsiij2}) we find
\begin{equation}
|\delta\psi\rangle=-2\pi i\lambda\int d\epsilon \hat{\cal N}^{\rm
 I}(\epsilon)\hat{\cal{N}}^{\rm II}(-\epsilon)|\psi^{\rm
 I}\psi^{\rm II}\rangle.
\end{equation} 
To summarize this section, we have written the {\em interacting} part of the outgoing wave function as the
charge fluctuation operators of the two dots, defined in Eq.\ (\ref{Nop}),
multiplying the {\em non-interacting} wave function $|\psi^{\rm I}\psi^{\rm II}\rangle$.

\section{Cross-correlation}

Having determined the wave function we are ready to calculate expectation values.
We are interested in an expectation value of the form $\hat{A}^{\rm I}\hat{B}^{\rm
  II}$. Here operator $\hat{A}^{\rm I}$ is a function of outgoing operators
$b^\dag$ and $b$, while $\hat{B}^{\rm II}$ depends on operators $d^\dag, d$. The
change in expectation value due to the interaction is given by
\begin{align}
\label{expvalue}
\langle \hat{A}^{\rm I}\hat{B}^{\rm II}\rangle-&\langle \hat{A}^{\rm
  I}\rangle_{\rm NI}\langle\hat{B}^{\rm II}\rangle_{\rm NI}
=\langle\psi^{\rm II}\psi^{\rm I}|\hat{A}^{\rm
  I}\hat{B}^{\rm II}|\delta\psi\rangle+{\rm H.C.}\nonumber\\
=&-2\pi i\lambda\int d \epsilon\left(
\langle\hat{A}^{\rm
  I}\hat{\cal N}^{\rm I}(\epsilon)\rangle_{\rm NI}\langle\hat{B}^{\rm
  II}\hat{\cal N}^{\rm
 II}(-\epsilon)\rangle_{\rm NI}\right.\nonumber\\
&\left.-\langle\hat{\cal N}^{\rm I}(\epsilon)\hat{A}^{\rm
  I}\rangle_{\rm NI}\langle\hat{\cal N}^{\rm
 II}(-\epsilon)\hat{B}^{\rm II}\rangle_{\rm NI}\right).
\end{align}
The subscript ${\rm NI}$ stands for non-interacting, hence $\langle A^{\rm
  I}\rangle_{\rm NI}=\langle\psi^{\rm I}| A^{\rm
  I}|\psi^{\rm I}\rangle$ and $\langle B^{\rm
  II}\rangle_{\rm NI}=\langle\psi^{\rm II}| B^{\rm
  II}|\psi^{\rm II}\rangle$, with $|\psi^{\rm I/II}\rangle$ the
  non-interacting wave functions of Eq.\ (\ref{defpsiout}).
We have written the expectation value of a certain operator $\hat{A}^{\rm
  I}\hat{B}^{\rm II}$ with respect to the interacting wave function as the
  expectation value of a new operator with respect to the non-interacting
  wave function. If we choose $\hat{A}^{\rm I}=\openone^{\rm I}$ and
  $\hat{B}^{\rm II}=\openone^{\rm II}$ we find that the interaction has no
  effect, i.e. the norm of the wave function is conserved up to linear order in
  $\lambda$, due to the unitarity of the two-particle scattering matrix.

In this paper we are interested in cross-correlations, i.e. one has
to substract 
the product $\langle\hat{A}^{\rm
  I}\rangle\langle\hat{B}^{\rm II}\rangle$. The expectation value
$\langle\hat{A}^{\rm I}\rangle$ can also be calculated with the help of Eq.\
(\ref{expvalue}), by taking the operator of the second system to be
the unity operator $\openone^{\rm II}$. We find
\begin{align}
\label{corr}
&\langle\delta \hat{A}^{\rm I}\delta \hat{B}^{\rm II}\rangle\equiv\langle \hat{A}^{\rm I}\hat{B}^{\rm II}\rangle-\langle
\hat{A}^{\rm I}\rangle\langle\hat{B}^{\rm II}\rangle\nonumber\\
&=\pi i\lambda\int d\epsilon \left(\langle[\hat{\cal N}^{\rm I}(\epsilon),\delta\hat{A}^{\rm
    I}]\rangle_{\rm NI}\langle\{\hat{\cal N}^{\rm
  II}(-\epsilon),\delta\hat{B}^{\rm II}\}\rangle_{\rm NI}+\right.\nonumber\\
&\qquad \left.\langle\{\hat{\cal N}^{\rm
  I}(\epsilon),\delta\hat{A}^{\rm I}\}\rangle_{\rm NI}\langle[\hat{\cal N}^{\rm
    II}(-\epsilon),\delta\hat{B}^{\rm II}]\rangle_{\rm NI}\right).
\end{align}
We have defined $\delta\hat{A}^{\rm I}=\hat{A}^{\rm
  I}-\langle\hat{A}^{\rm I}\rangle$, $\delta\hat{B}^{\rm II}=\hat{B}^{\rm II}-\langle\hat{B}^{\rm II}\rangle$.
Eq.\ (\ref{corr}) shows that to calculate correlations, one has to calculate the
  commutator $[,]$ and anti-commutator $\{,\}$ of the operator of interest and the
  charge fluctuation operator.

To summarize, we have calculated expectation values of non-interacting
operators with respect to an interacting wave function. The final result Eq.\
(\ref{corr}) however contains operators modified by the interaction and
non-interacting wave functions. This suggests that the effects of the
interaction can be completely incorporated into the operators. We define the
effective operators
\begin{align}
\label{eff}
&&\hat{A}^{\rm I}_{\rm eff}&=\hat{A}^{\rm I}+\lambda \int d \epsilon \langle [2\pi i\hat{\cal N}^{\rm I}(\epsilon),\hat{A}^{\rm I}]\rangle_{\rm NI}\, \hat{\cal N}^{\rm II}(-\epsilon).  \\
&&\hat{B}^{\rm II}_{\rm eff}&=\hat{B}^{\rm II}+\lambda \int d \epsilon \langle [2\pi i\hat{\cal N}^{\rm II}(-\epsilon),\hat{B}^{\rm II}]\rangle_{\rm NI}\, \hat{\cal N}^{\rm I}(\epsilon).  
\end{align}
The first part of these equations is just the non-interacting operator while the second part
describes the effect of the interaction.
These effective operators give exactly the same result as Eq.\ (\ref{corr}), when evaluated with respect to the non-interacting
wave function, i.e. 
\begin{align}
\frac{1}{2}\langle\delta\hat{A}^{\rm I}_{\rm eff} \delta\hat{B}^{\rm II}_{\rm eff}+\delta\hat{B}^{\rm II}_{\rm eff} \delta\hat{A}^{\rm I}_{\rm eff}\rangle_{\rm NI}=\langle\delta \hat{A}^{\rm I}\delta \hat{B}^{\rm II}\rangle.
\end{align}
We have again defined $\delta\hat{A}^{\rm I}_{\rm eff}=\hat{A}^{\rm I}_{\rm eff}-\langle\hat{A}^{\rm I}_{\rm eff}\rangle$.
The existence of cross-correlations can now be understood in the following
way. Due to the interaction, an operator on dot ${\rm I}$ depends on the
charge fluctuation operator on dot ${\rm II}$ and vice-versa. If the charge fluctuations on
dot ${\rm II}$ are correlated with the fluctuations in the non-interacting
operator $\hat{B}^{\rm II}$, or if the charge fluctuations in system ${\rm I}$
correlate with fluctuations in $\hat{A}^{\rm I}$, we find a cross-correlation
(in linear order in $\lambda$).
In the next few paragraphs we will consider current cross-correlations and
we will find that the effective operators are very convenient to express the cross-correlation and directly point to the origin of the correlations.

\section{Zero-frequency current cross-correlation}

The current operator in the left lead of dot ${\rm I}$ is given by \cite{Bla00}
\begin{align}
\hat{I}^{\rm I}_{L}(\omega)&\equiv \hat{I}^{\rm I, in}(\omega)-\hat{I}^{\rm I, out}(\omega),\\
\hat{I}_L^{\rm I,in}(\omega)&=e\sum_{i=1}^{N_L^{\rm I}}\int dE\,\,a_{i}^\dag(E+\hbar\omega) a_{i}(E),\\
\hat{I}_L^{\rm I,out}(\omega)&=e\sum_{i=1}^{N_L^{\rm I}}\int dE\,\,b_{i}^\dag(E+\hbar\omega) b_{i}(E).
\end{align}
In an equivalent way a current can be defined for the second dot.
We are interested in the zero-frequency current cross-correlation \cite{Bla00}
\begin{equation}
\label{PI1I2def}
P_{I_1I_2}(\omega)2\pi\delta(\omega+\omega')=
\langle \delta\hat{I}_L^{\rm I}(\omega)\delta\hat{I}_L^{\rm
  II}(\omega')+\delta\hat{I}_L^{\rm II}(\omega') \delta\hat{I}_L^{\rm
  I}(\omega)\rangle.
  \end{equation}
Eqs.\ (\ref{corr}) and (\ref{eff})
describe cross-correlations of {\em outgoing} operators only and can be used
to calculate the cross-correlation between outgoing currents.
Since $a_l^\dag(E)a_l(E)|\psi\rangle=\langle
a_l^\dag(E)a_l(E)\rangle|\psi\rangle$ (cf. to Eq.\ (\ref{psiin})), $\hat{I}^{\rm
  I,in}$ will not contribute to the cross-correlations, the same
is true for the incoming current on dot ${\rm II}$.

We calculate the effective operators of Eq.\ (\ref{eff}) with  $\hat{A}^{\rm I}=\hat{I}^{\rm I,out}(\omega)$ and
$\hat{B}^{\rm II}=\hat{I}^{\rm II,out}(\omega')$.
We will need the commutator
\begin{align}
\langle[\hat{\cal N}^{\rm I}(\epsilon), \hat{I}_L^{\rm I, out}(\omega)]\rangle_{\rm
  NI}=&
-  V^{\rm I}{\delta(\hbar\omega-\epsilon)}\nonumber\\
&\frac{G^{\rm I}(E_F+\hbar\omega)-G^{\rm I}(E_F)}{i \hbar\omega},
\end{align}
where $G^{\rm I}(E_F)=\frac{e^2}{2\pi\hbar}Tr[S^{\rm I\dag}_{RL}(E_F)S^{\rm I}_{RL}(E_F)]$ is the
  non-interacting conductance \cite{Bla00}. The element
  $S_{RL}^{\rm I}$ is the block of the scattering matrix connecting channels in the right
  and left lead. 
We find effective current operators
\begin{align}
\label{Ieff0}
\hat{I}^{\rm I,out}_{L,\rm eff}(\omega)=&\langle\hat{I}_L^{\rm
  I,out}(\omega)\rangle_{\rm NI}+\delta\hat{I}_L^{\rm I,out}(\omega)-
2\pi\lambda V^{\rm I}\times\nonumber\\
&\frac{ G^{\rm I}(E_F+\hbar\omega)-G^{\rm I}(E_F)}{\hbar\omega}\hat{\cal N}^{\rm II}(-\hbar\omega),\\
\label{Ieffom}
\hat{I}^{\rm II,out}_{L,\rm eff}(\omega')=&
\langle\hat{I}_L^{\rm II,out}(\omega')\rangle_{\rm NI}+\delta\hat{I}_L^{\rm II,out}(\omega')-2\pi\lambda
 V^{\rm II}\times\nonumber\\
&\frac{G^{\rm II}(E_F+\hbar\omega')-G^{\rm II}(E_F)}{\hbar\omega'}\hat{\cal N}^{\rm I}(-\hbar\omega').
\end{align}  
These expressions for the effective current operator are very intuitive: due to the interaction a
fluctuation in the number of charges on dot ${\rm I}$ (${\rm II}$) causes a fluctuating
potential energy for charges entering the other dot proportional to $\lambda \hat{\cal
  N}^{\rm I}$ ($\lambda\hat{\cal
  N}^{\rm II}$) and if the conductance is energy dependent a charge
fluctuation on one dot will lead to a current fluctuation on the other dot. 

The cross-correlation linear in  $\lambda$ is due to the correlation of the charge
 fluctuations and
 the non-interacting current fluctuation,
\begin{align}
\label{corrtotal}
P_{I_1I_2}(\omega)\delta(\omega+\omega')=&\frac{1}{2\pi}\langle\{\delta\hat{I}^{\rm
    I,out}_{\rm eff}(\omega),\delta\hat{I}^{\rm
    II,out}_{\rm eff}(\omega')\}\rangle_{\rm NI}=\nonumber\\
&\lambda V^{\rm II}\frac{G^{\rm
    II}(E_F+\hbar\omega')-G^{\rm II}(E_F)}{-\hbar\omega'}\times\nonumber\\
&\langle\{\hat{N}^{\rm
    I}(-\hbar\omega'),\delta\hat{I}^{\rm I,out}_L(\omega)\}\rangle_{\rm
    NI}\nonumber\\
&+\lambda V^{\rm I}\frac{G^{\rm
    I}(E_F+\hbar\omega)-G^{\rm I}(E_F)}{-\hbar\omega}\times\nonumber\\
&\langle\{\hat{N}^{\rm
    II}(-\hbar\omega),\delta\hat{I}^{\rm II,out}_L(\omega')\}\rangle_{\rm NI}.
\end{align}
We calculate
\begin{align}
\label{term2}
&\langle\{\hat{\cal N}^{\rm I}(-\hbar\omega'),\delta\hat{I}^{\rm I, out}(\omega)\}\rangle_{\rm
  NI}=-\frac{e}{\hbar}\delta(\omega+\omega')\sum_{i=1}^{N_L^{\rm I}}\nonumber\\
&\sum_{j}^{\rm
  N^{\rm I}}\sum_{k}^{\rm
  N^{\rm I}}\int dE\left[S^{\rm I}_{ij}(E-\hbar\omega')N^{\rm
  I}_{jk}(E-\hbar\omega',E)S^{\rm I\dag}_{ik}(E)\right]\nonumber\\
&\left(f^{\rm I}_j(E-\hbar\omega')(1-f^{\rm I}_k(E))+f^{\rm I}_k(E)(1-f^{\rm
  I}_j(E-\hbar\omega')\right).
\end{align}
Here $f_i^{\rm I}=\lim_{T\rightarrow 0}\left[(1+\exp[(E-E_F-eV^{\rm I|}_i)/k_B T])^{-1}\right]$ is the zero temperature Fermi-function at mode $i$ of dot
${\rm I}$.

Combining Eqs.\ (\ref{corrtotal}) and (\ref{term2})  
we can write the zero-frequency current cross-correlation in terms of the
non-interacting scattering matrices.  Since
we assume zero temperature we have to evaluate all quantities at the Fermi energy and
we suppress the energy arguments. 
We find
\begin{align}
\label{PI1I2}
&P_{I_1I_2}(0)\equiv P_{I_1I_2}=\nonumber\\
&\frac{\lambda e^2 } {2\hbar}  \left(\frac{\partial G^{\rm I}}{\partial
  E_F}V^{\rm I} {\rm Tr}[S^{\rm II}_{LL}N^{\rm II}_{LR}S_{LR}^{\rm
  II\dag}+
S^{\rm II}_{LR}N^{\rm II}_{RL}S^{\rm II\dag}_{LL}]|V^{\rm II}|  
\right. \nonumber \\ 
&\left. +\frac{\partial G^{\rm II}}{\partial
  E_F}V^{\rm II}
  {\rm Tr}[S^{\rm I}_{LL}N^{\rm I}_{LR}S_{LR}^{\rm
  I\dag}+S^{\rm I}_{LR}N^{\rm I}_{RL}S^{\rm I\dag}_{LL}] |V^{\rm I}|\right).
\end{align}

\section{Chaotic quantum dots}

{\begin{figure}[t]
\begin{center}
\includegraphics[width=8cm]{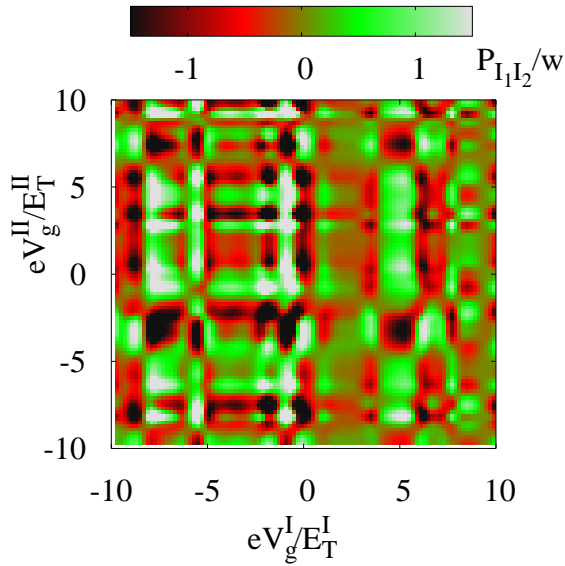}
\end{center}
\caption{(color online) The current cross-correlation $P_{I_1I_2}$ for dots with scattering matrices
  $S^{\rm I}(E+eV^{\rm I}_g)$ and $S^{\rm II}(E+V_g^{\rm II})$ for different
  gate voltages. The scattering matrices are calculated from numerically
  generated $M\times M$ random Hamiltonians (of the Gaussian Unitary Ensemble) and Eq.\ (\ref{SvsH}). The value
  of $P_{I_1I_2}$, in units of the root mean squared $w$ of the
  fluctuations, is represented by its color (gray
  scale), indicated by the scale at the top. 
  The gate voltage is in units of
  the Thouless energy $E_T^{\rm I/II}$.
 We chose $N^{\rm I}=N^{\rm II}=8$,
  $N_L^{\rm I/II}=N_R^{\rm I/II}=\frac{1}{2}N^{\rm I/II}$, $M=200$.}
\label{figcorr}
\end{figure}

While the previous derivation is valid for any type of scatterer, we will now
assume specific systems: two chaotic quantum dots. We use Random Matrix
Theory (RMT) to calculate mesoscopic averages \cite{Bro96,waves} by integrating
the scattering matrices over the Circular Unitary Ensemble (CUE). To
distinguish the ensemble average from the expectation values, we denote
it by $\langle\langle\hdots\rangle\rangle$. 
The ensemble averaged quantities $\langle\langle\frac{\partial G}{\partial
  E}\rangle\rangle$ and $\langle\langle{\rm Tr}[S_{LL}N_{LR}S_{LR}^\dag+S_{LR}N_{RL}S^{\dag}_{LL}]\rangle\rangle$ are both zero.  
The cross-correlation will fluctuate from ensemble member to ensemble member and can be either negative
or positive but is zero on average.

The behavior of $P_{I_1I_2}$ is illustrated in Fig.\
\ref{figcorr}. 
To imitate a real experiment we assume that the variation in scattering
matrices is due to a change in the gate voltages $V_g^{\rm I/II}$, which enter
the scattering matrix via the energy argument; $S^{\rm I/II}(E,V_g^{\rm I/II})=
S^{\rm I/II}(E+eV_g^{\rm I/II})$.
To numerically evaluate scattering matrices at different energies, we start from two $M\times M$ random Hamiltonians
$H^{\rm I}$, $H^{\rm II}$, with the distribution of the Gaussian Unitary Ensemble (GUE). The scattering
matrix at energy $E$ is related to the Hamiltonian via the relation \cite{Bee97}
\begin{align}
\label{SvsH}
S^{\rm I/II}(E)=&1-2\pi i\times \nonumber\\&W^{\rm I/II}\frac{1}{E-H^{\rm I/II}+i\pi W^{\rm
    I/II\dag}W^{\rm I/II}}W^{\rm I/II\dag}.
\end{align}
Here $W^{\rm I/II}$ is an $N^{\rm I/II}\times M$ matrix describing the coupling between lead and
dot. For ballistic leads the combination $W^{\rm I/II}W^{\rm I/II\dag}$ has eigenvalues
${M\delta^{\rm I/II}}/{\pi^2}$, where $\delta^{\rm I/II} $ is the level spacing in the dot {\rm I/II}. We assume non-zero matrix elements
$W^{\rm I/II}_{ij}=\delta_{ij} {\sqrt{M\delta^{\rm I/II}}}/{\pi}$ for $i=1\hdots
N^{\rm I/II}$. 
With these definitions the cross-correlation of Eq.\ (\ref{PI1I2}) can be calculated
for different gate voltages and the result is
indicated by the color (gray scale) in Fig.\ \ref{figcorr}. We find
fluctuations that are smooth up to a scale set by the Thouless energy of the open
dot, $E_T^{\rm I/II}=N^{\rm I/II}\delta^{\rm I/II}/2\pi$, suggesting that two
scattering matrices with different energies are truly independent if the
energy difference exceeds $E_T$. This is indeed a well-known property of
mesoscopic systems \cite{Imr97}.
The magnitude of the
fluctuations is set by $w$ which we will proceed to calculate.

To determine the magnitude of the fluctuations we calculate (in leading order in $N=N_L+N_R$)
\begin{align}
&\langle\langle \left(\frac{\partial G}{\partial
  E_F}\right)^2\rangle\rangle=\left(\frac{e^2}{\hbar\delta}\right)^2\frac{2N^2_LN^2_R}{N^6},\\
&\langle\langle{\rm Tr}^2[S_{LL}N_{LR}S_{LR}^{\dag}+S_{LR}N_{RL}S^{\dag}_{LL}]\rangle\rangle=
\frac{2N_L^2N_R^2}{ N^6\delta^2},
\\
&\langle\langle{\rm Tr}[S_{LL}N_{LR}S_{LR}^{\dag}+S_{LR}N_{RL}S^{\dag}_{LL}]\frac{\partial G}{\partial
  E_F}\rangle\rangle=0.
\end{align}
We have suppressed the superscript ${\rm I/II}$ because the ensemble averaged result is the
same for both dots.
Putting everything together we find for the magnitude of the fluctuations
\begin{align}
\label{rms}
w^2\equiv\langle\langle P_{I_1I_2}^2\rangle\rangle=\left(\frac{\lambda e^4 V^{\rm I}V^{\rm II}}{\hbar^2\delta^{\rm I}\delta^{\rm
    II}}\right)^2\frac{2(N_L^{\rm I}N_R^{\rm I}N_L^{\rm II}N_R^{\rm
    II})^2}{(N^{\rm I}N^{\rm II})^6}.\,\,\,
\end{align}
The magnitude of the fluctuations decreases as $1/(N^{\rm I}N^{\rm II})^2$ and
is maximal for symmetric dots. A similar dependence of $1/N^4$ was found for
the fluctuations of the magnetic-field asymmetry of the nonlinear conductance
of a mesoscopic conductor \cite{San04,Pol06}.

In the experiment of Ref.\ \cite{McC07} the zero-frequency cross-correlation of two coupled
quantum dots is also shown as a function of gate voltages, similar to Fig.\
\ref{figcorr}. The cross-correlation can also be positive or negative, but
does not
exhibit the random fluctuations illustrated in Fig. \ref{figcorr}. In
contrast, planes with negative and
positive sign are ordered in a checkerboard pattern.
In the
experiment the leads have tunnel barriers and transport is predominantly through one level
in the quantum dot. The energy scale of the interaction is not small with
respect to the applied voltage and we can therefore not apply our theory to
that particular set-up. In this paper we have studied the cross-correlation in
another regime where we find a different result. 

\section{Discussion and conclusion}
We would like to comment on the validity of our approach. 
We have only considered two-particle interactions between particles in different
conductors and we have neglected the intra-dot interactions. Since we are
interested in cross-correlations, this is certainly a good approximation if
both the inter-dot and intra-dot interactions are small. In that case a
perturbative approach in both types of interactions is sufficient, and only
the leading order result in inter-dot interactions will contribute to the
cross-correlations. However, the intradot interaction is typically not small. 
To leading order in the inverse 
number of channels \cite{Bro05}, in the open quantum dots considered here, the interaction inside a dot is well described by a Hartree potential which can be incorporated into an effective scattering matrix.

To conclude we have derived the multi-particle outgoing wave function for two weakly coupled
conductors and used it to calculate cross-correlations for transport
experiments. We have defined effective operators which give the same result
when evaluated with respect to the non-interacting wave function.
We have
specifically focused on the zero-frequency current cross-correlation of two chaotic quantum dots. This correlation fluctuates from sample-to-sample and is zero on average. We have quantified the magnitude of the fluctuations and verified the behavior numerically. The increasing sensitivity of noise measurements should bring a detailed investigation of the current correlations discussed here within experimental reach.

\section*{Acknowledgements}
The work was supported by the Swiss National Science Foundation and the EU Marie Curie RTN "Fundamentals of Nanoelectronics", MCRTN-CT-2003-504574.


\begin{thebibliography}{99}

\bibitem{Goo07}  M.\ C.\ Goorden and M.\ B\"uttiker, 
                 Phys.\ Rev.\ Lett.\ {\bf 99}, 146801 (2007). 

\bibitem{McC07}  D.\ T.\ McClure, L.\ DiCarlo, Y.\ Zhang, H.-A.\ Engel, C.\ M.\                                    Marcus,M.\ P.\ Hanson, and A.\ C.\  Gossard, 
                 Phys.\ Rev.\ Lett.\ {\bf98}, 056801  (2007).

\bibitem{Bro05}  P.\ W.\ Brouwer, A.\ Lamacraft, and K.\ Flensberg,
                 Phys.\ Rev.\ B {\bf 72}, 075316 (2005).



\bibitem{Mor01}  N.\ A.\ Mortensen, K.\ Flensberg, and A.~-P.\ Jauho,
                 Phys.\ Rev.\ Lett.\ {\bf 86}, 1841 (2001).
                 
\bibitem{Yama}   M.
 Yamamoto, M.\ Stopa, Y.\ Tokura, Y.\
		             Hirayama, and S.\ Tarucha, Science {\bf 313}, 204 (2006).

\bibitem{moty1}  D.\ Rohrlich, O.\ Zarchin, M.\ Heiblum, D.\ Mahalu, and V.\ Umansky, 
                 Phys.\ Rev.\ Lett.\ {\bf 98}, 096803 (2007). 
                

\bibitem{Ji03}   Y.\ Ji, Y.\ Chung, D.\ Sprinzak, M.\ Heiblum, D.
 Mahalu,
                 and H.\ Shtrikman, Nature {\bf 422}, 415 (2003).  

\bibitem{litv}   L.\ V.\ Litvin, H.-P.
 Tranitz, W.\ Wegscheider, and C.\ Strunk,
                 Phys.\ Rev.\ B {\bf 75}, 033315 (2007). 

\bibitem{roche1} P.\ Roulleau, F.\ Portier, D.\ C.\ Glattli, P.\ Roche, A.\ Cavanna, G.\ Faini, U.\ Gennser, 
                 and D.\ Mailly, Phys.\ Rev.\ B {\bf 76}, 161309(R) (2007).
                 
\bibitem{roche2} P.\ Roulleau, F.\ Portier, P.\ Roche, A.\ Cavanna, G.\ Faini, U.\ Gennser, and D.\ Mailly,  Phys.\ Rev.\ Lett.\ {\bf 100}, 126802 (2008).
               

\bibitem{roche3} P.\ Roulleau, F.\ Portier, P.\ Roche, A.\ Cavanna, G.\ Faini, U.\ Gennser, and D.\ Mailly
                 (unpublished). arXiv:0802.2219 
                 
\bibitem{litv2}  L. V. Litvin, A. Helzel, H.-P. Tranitz, W. Wegscheider, and C. Strunk    
                 (unpublished). arXiv:0802.1164   




\bibitem{See}    G.\ Seelig and M.\ B\"uttiker, Phys.\ Rev.\ B {\bf 64}, 245313 (2001).

\bibitem{Chu}    V.S.-W.\ Chung, P.\ Samuelsson, and M.\ B\" uttiker, Phys.\ Rev.\ B {\bf 72}, 125320 (2005).


\bibitem{moty3}  I.\ Neder, F.\ Marquardt, M.\ Heiblum, D.\ Mahalu, and V.\ Umansky,
                 Nature Physics {\bf 3}, 534 (2007).              
                 
                


\bibitem{Ned07}  I.\ Neder and F.\ Marquardt, New Journal of Physics {\bf 9},
                 112 (2007).

\bibitem{ned08}   I.\ Neder and E.\ Ginossar, Phys.\ Rev.\ Lett.\ {\bf 100}, 196806 (2008).
                 
\bibitem{sim07}  S.-C.\ Youn, H.-W.\ Lee, and H.\ -S.\ Sim, Phys.\ Rev.\ Lett.\ {\bf 100}, 196807 (2008).
                 

\bibitem{sukh08} I. P. Levkivskyi and E. V. Sukhorukov (unpublished).  
                 arXiv:0801.2338

\bibitem{gabe}   J.\ Gabelli, G.\ F\`eve, J.-M.\ Berroir, B.\ Pla\c{c}ais, A.\ Cavanna, B.\ Etienne, 
                 Y.\ Jin, and D.\ C.\ Glattli, Science {\bf 313}, 499 (2006).



\bibitem{nigg}   S.\ E.\ Nigg, R.\ L\'opez and M.\ B\"{u}ttiker,
                 Phys. Rev. Lett. {\bf 97}, 206804 (2006); 
                 M.\ B\" {u}ttiker and S.\ E.\ Nigg, Nanotechnology {\bf 18}, 044029 (2007);
                 S.\ E.\ Nigg and M.\ B\"{u}ttiker, Phys.\ Rev.\ B {\bf 77}, 085312 (2008).
                  
\bibitem{feve}   G.\ F\`{e}ve, A.\ Mah\'e, J.-M.\ Berroir, T.\ Kontos, B.\ Pla\c{c}ais, D.\ C.\                           Glattli, A.\ Cavanna, B.\ Etienne, and Y.\ Jin,
                 Science {\bf 316}, 1169 (2007).  
                  

\bibitem{moska}  M.\ Moskalets, P.\ Samuelsson, and M.\ B\"{u}ttiker, Phys.\ Rev.\ Lett.\ {\bf 100},                       086601 (2008).

\bibitem{qing03} Q.-F.\ Sun, H.\ Guo, and J.\ Wang, 
                 Phys.\ Rev.\ B {\bf 68}, 035318 (2003).

\bibitem{PRL92}  M. B\"{u}ttiker, Phys.\ Rev.\ Lett.\ {\bf 68}, 843 (1992).

\bibitem{Sam04} P.\ Samuelsson, E.\ V.\ Sukhorukov, and M.\ B\"{u}ttiker,
                Phys.\ Rev.\ Lett.\ {\bf 92}, 026805 (2004).
  
\bibitem{Ned07exp} 	 I.\ Neder, N.\ Ofek, Y.\ Chung, M.\ Heiblum, D.\ Mahalu, and V.\ Umansky,
                     Nature {\bf 448}, 333 (2007).

\bibitem{LL93}   L.\ S.\ Levitov and G.\ B.\ Lesovik, JETP Lett. {\bf 58}, 230 (1993). 

\bibitem{Sim05}  H.-S.\ Sim and E.\ V.\ Sukhorukov, 
                 Phys.\ Rev.\ Lett.\ {\bf 96}, 020407 (2006).  

\bibitem{Jac97}  Ph.\ Jacquod and D.\ L.\ Shepelyansky, Phys.\ Rev.\ Lett.\     {\bf 78}, 4986 (1997).  
                 
\bibitem{Leb07}  A.\ V.\ Lebedev, G.\ B.\ Lesovik, and G.\ Blatter (unpublished). arXiv:0711.4308

\bibitem{dhar08} A.\ Dhar, D.\ Sen, and D.\ Roy (unpublished). arXiv:0802.2380





\bibitem{Sel06}  E.\ Sela, Y.\ Oreg, F.\ von Oppen, and J.\ Koch, Phys.\ Rev.\ Lett.\ {\bf 97}, 086601 (2006).

\bibitem{moty4}  O.\ Zarchin, M.\ Zaffalon, M.\ Heiblum, D.\ Mahalu, and V.\ Umansky (unpublished). arXiv:0711.4552

\bibitem{hur7}  P. Vitushinsky, A. A. Clerk, and K. Le Hur, Phys. Rev. Lett. {\bf 100}, 036603 (2008).
  
\bibitem{Ped98}  M.\ H.\ Pedersen, S.\ A.\ van Langen, and M.\ B\"{u}ttiker,
                 Phys.\ Rev.\ B {\bf 57}, 1838 (1998).
                 
                 
\bibitem{But99}  M.\ B\"uttiker, J.\ Math.\ Phys.\, {\bf 37}, 4793 (1996).


\bibitem{Smi60}  F.\ T.\ Smith, Phys.\ Rev.\ {\bf 118}, 349 (1960).  

\bibitem{Bro97}  P.\ W.\ Brouwer, S.\ A.\ van Langen, K.\ M.\ Frahm, M.\ B\"uttiker, and C.\ W.\ J.\ Beenakker, Phys.\ Rev.\ Lett.\ {\bf 79}, 913 (1997).
  


\bibitem{Ros65}  L.\ Rosenberg, Phys.\ Rev.\ {\bf 140}, B217 (1965).                                          
                  
                  

\bibitem{Sam03}   P.\ Samuelsson, E.\ V.\ Sukhorukov, and M.\ B\"uttiker, Phys.\
                  Rev.\ Lett.\ {\bf 91}, 157002 (2003).

\bibitem{Sam05}   P.\ Samuelsson,  E.\ V.\ Sukhorukov, and M.\ B\"uttiker, New.\
                  J.\ Phys.\ {\bf 7}, 176 (2005).
  
\bibitem{Bla00}   Y.\ M.\ Blanter and M.\ B\"{u}ttiker, Phys.\ Rep.\ {\bf 336}, 1 (2000).

\bibitem{Bro96}   P.\ W.\ Brouwer and C.\ W.\  J.\ Beenakker,
                  J.\ Math.\ Phys.\ {\bf 37}, 4904 (1996). 

\bibitem{waves}   P.\ W.\ Brouwer, K.\ M.\ Frahm, and C.\ W.\ J.\ Beenakker,
                  Waves~in~Random~Media {\bf 9}, 91 (1999).

\bibitem{Bee97}   C.\ W.\ J.\ Beenakker, Rev.\ Mod.\ Phys.\ {\bf 69}, 731 (1997).

\bibitem{Imr97}   Y.\ Imry, {\em Introduction to mesoscopic physics}, (Oxford
                  University Press, Oxford, 1997).
                  
\bibitem{San04}   D.\ S\'anchez and M.\ B\"uttiker, 
                  Phys.\ Rev.\ Lett.\ {\bf 93}, 106802 (2004).
                  
\bibitem{Pol06}   M.\ L.\ Polianski and M.\ B\"uttiker,  
                  Phys.\ Rev.\ Lett.\ {\bf 96}, 156804 (2006).                   

\end{thebibliography}
\end{document}